\documentstyle[12pt,epsf]{article}
\def\beq{\begin{eqnarray}}    
\def\eeq{\end{eqnarray}}      

\def\al{\alpha}
\def\be{\beta}

\def\ga{\gamma}
\def\de{\delta}

\def\ka{\kappa}
\def\la{\lambda}
\def\na{\nabla}
\def\pa{\partial}

\def\si{\sigma}

\def\ph{\varphi}

\def\Ga{\Gamma}
\def\De{\Delta}

\begin{document}

\hfill gr-qc/9810032

\hfill  8 of October, 1998

\vskip 3mm

\begin{center}

{\Large\sl  
Anomaly-induced effective action for gravity and inflation}
\vskip 8mm

{\large  J.C. Fabris $^a$}
 \footnote{Electronic address: fabris@cce.ufes.br},$\,$
{\large A.M. Pelinson $^b$} 
 \footnote{Electronic address: ana@fisica.ufjf.br},$\,$
{\large I.L. Shapiro $^{b,c}$}
 \footnote{Electronic address: shapiro@fisica.ufjf.br},$\,$
\vskip 3mm

a. {\sl Departamento de F\'{\i}sica -- CCE, 
Universidade Federal de Espir\'{\i}to Santo, ES, Brazil}
\vskip 2mm

b. {\sl Departamento de F\'{\i}sica -- ICE, 
Universidade Federal de Juiz de Fora, MG, Brazil}
\vskip 2mm

c. {\sl Tomsk State Pedagogical University, Tomsk, Russia}

\end{center}

\vskip 8mm

\noindent
{\large\it Abstract}.$\,\,$
In the very early Universe the matter may be described by the free 
radiation, that is by the set of massless fields with negligible
interactions between them. Then the dominating quantum effect
is the trace anomaly which comes from the renormalization of
the conformal invariant part of the vacuum action. The anomaly-induced
effective action can be found with accuracy to an arbitrary conformal
functional which vanishes for the special case of the
conformally flat metric. This gives the solid basis for the study
of the conformally-flat cosmological solutions, first of which was
discovered by Mamaev and Mostepanenko and by Starobinski in 1980. 
Treating the anomaly-induced action as quantum correction to the
Einstein-Hilbert term we explore the possibility to have inflationary 
solutions, investigate their dependence on the initial data and 
discuss the restrictions in considering the density perturbations. 
The shape of inflationary solutions strongly depends on the underlying 
gauge model of the elementary particles physics. Two special 
cases are considered: Minimal Standard Model and the matter sector 
of $N=8,\,D=4$ supergravity. It turns out that inflation is almost 
inevitable consequence of the great difference between Planck mass 
and the mass of the heaviest massive particle.

\section{Introduction.}

The standard cosmological model \cite{frid} was constructed soon
after the creation of General Relativity and it describes the
present stage of the Universe pretty well (see, for example,
\cite{cosm1}). At the same time this model fails to explain the
early period of the evolution, and hence some new theoretical idea
is required. It is commonly accepted that problems of
initial singularity, horizon, flatness and monopoles
may be solved if one supposes that the early Universe passed through
the period of a very fast (more than the speed of light) inflation
when its size increased in more than 60 orders. Moreover, the origin
of cosmological perturbations has an elegant explanation within
such scenario. On the other hand,
the source of inflation still remains unclear. In the first works
devoted to the exponential expansion of the Universe the origin of 
it was considered in the quantum effects of vacuum \cite{mamo,star}
(see also \cite{fhh}),
but the real explosion of interest to it started after the paper
\cite{guth} where inflation was induced as a consequence of the 
Spontaneous Symmetry Breaking in the Standard Model of
electroweak interactions. The further development has shown that
the last cosmological scenario was not perfect and
many other models, theories and approaches were considered.
Finally it was realized that all the requirements to the inflationary
model can be satisfied only if one introduces some new scalar
field called inflaton, which has a potential of a very special form
(see, for example, \cite{review1,review2,review3} for a comprehensive 
review). 

Here we intend to go backward to the historic origin of inflation 
\cite{mamo,star} and
derive it from the quantum effects of pure radiation on
curved background. It will be shown that the inflationary solution
doesn't require any fine-tuning or special choice of the parameters
of the underlying conformal invariant matter fields but is absolutely
natural and exists for a wide set of initial data. The exit from the
inflationary phase may be also naturally explained by the nonstability
of the exponential solution with respect to the density perturbations
that can be relevant at the end of inflationary epoch.
In this letter our
consideration is restricted by the conformally flat metrics, but  
this does not mean that the similar results can not be achieved
for the $\,k=+1\,$ or $\,k=-1\,$ cases.

The paper is organized as follows. In the next section we consider
the model for the very Early Universe filled by the free radiation. 
This radiation is composed by the
set of free, massless, conformally invariant matter fields on an
arbitrary curved background. We review the derivation of the 
quantum correction to the classical action of gravity 
\cite{rei,frts}. This correction results from the trace anomaly 
\cite{duff} and it contains an arbitrary conformal
invariant functional. However, for the conformally
flat background metric this functional is irrelevant and hence in the
framework of the above model the solution of \cite{rei,frts} is exact.
In section 3 the particular inflationary solution is presented and 
its nonstability is established. Section 4 contains the results of  
numerical investigation and the physical analysis of the inflation. 
In the last section we draw some conclusions and outline the prospects 
for further studies.

\section{Free fields and quantum corrections}

The action of the ideal liquid filling the Universe can be written as
\beq
S_{M}=
\int d^{4}x\sqrt{-g}\left\{\,- \frac{1}{2}\left( p+\rho \right) 
u^{\mu}u^{\nu }g_{\mu \nu }+\frac{3}{2}p+\frac{1}{2}\rho \right\}\,. 
\label{liq}
\eeq
It is easy to see that for the conformally flat cosmological metric
the variation of 
this action with respect to the conformal factor of the metric
$$
\frac{\de S_M}{\de a} = (3p - \rho)\,a^3\, V
$$
vanishes for $p = \frac{\rho}{3}$, and therefore $\,T^\mu_\mu = 0\,$, 
so the radiation 
should be described by the conformal invariant field actions.

In the framework of the asymptotically free gauge theories the 
interactions between matter fields are weakened in the high energy 
limit. Therefore we start from the action of free massless conformal 
invariant fields: $N_0$ scalars (spin-0), 
$N_{1/2}$ spinors (Dirac, spin-1/2) and $N_1$ abelian vectors 
(spin-1). All $N$'s indicate a number of fields (not multiplets) 
in curved space-time, taking conformal coupling with scalar curvature
for the scalars.
\beq
S_{matter} = 
\int d^4 x\sqrt{-g}\left\{\sum_{i=1}^{N_0} 
\frac12(\na^\mu\ph_i\na_\mu\ph_i +\frac{1}{6}R\ph^2)
+ i \sum_{j=1}^{N_{1/2}} {\bar \psi}_j\ga^{\mu}\na_\mu \psi_j
 - \frac14\sum_{k=1}^{N_1} F_{k,\mu\nu}{F_k}^{\mu\nu}
\right\}\,.
\label{fields}
\eeq

For the free fields on curved background the only possible divergences 
are the one-loop vacuum ones. Therefore the renormalizability requires
the action of vacuum to be introduced in the form (see \cite{birdav}
and \cite{book} for the introduction)
\beq
S_{vacuum} = \int d^4 x\sqrt{- g}\,\left\{l_1C^2 + l_2E + l_3{\Box}R
\right\}
\label{vacuum}
\eeq
where $l_{1,2,3}$ are some new (with respect to the flat space-time)
parameters, $C^2$ is square of the Weyl tensor and $E$ is the
integrand of the Gauss-Bonnet topological (in $d=4$) invariant.
The vacuum action (\ref{vacuum}) includes only conformal invariant 
and surface terms. For the convenience of the reader we reproduce 
known formulas
$$
R_{\mu\nu\al\be}^2 = 2C^2-E+\frac13\, R^2
\,,\,\,\,\,\,\,\,\,\,\,
R_{\mu\nu}^2 = \frac12\,(C^2-E)+\frac13 R^2
$$
One has to notice that the introduction of 
conformal non-invariant terms like $\,\int\sqrt{-g}R\,$ or
$\,\int\sqrt{-g}R^2\,$ is possible but not necessary for the 
renormalization of the free conformal invariant theories. 
The renormalization of vacuum leads to the conformal 
(trace) anomaly \cite{duff} which enables one to 
find the nonconformal 
part of the effective action of vacuum \cite{rei,frts}. 
For the sake of 
completeness and also to fix notations and signs we reproduce the
main steps of this derivation.

The form of the one-loop divergences is well-known 
(see \cite{birdav,book}).
\beq
{\bar \Ga_{div}}^{(1)}  = 
 - \,\frac{\mu^{D-4}}{(D-4)}\,
\int d^D x\sqrt{-g}\,\{ \,w C^2 + b E + c {\Box} R\,\}\,. 
\label{divs}
\eeq
Here $\,D\,$
is the parameter of dimensional 
regularization and\footnote{We use notation ${R^\al}_{\be\ga\de}=
\pa_\ga \Ga^\al_{\be\de} - ...\,$.}
$$ 
w = \frac{1}{(4\pi)^2}\,\left(\,
 \frac{1}{120}\,N_0 + \frac{1}{20}\,N_{1/2} + 
\frac{1}{10}\,N_1\,\right)\,,
$$$$
b\, =\, -\,\frac{1}{(4\pi)^2}\,\left(\, \frac{1}{360}\,N_0 + 
\frac{11}{360}\,N_{1/2} + \frac{31}{180}\,N_1 \,\right)\,,
$$
\beq
c = \frac{1}{(4\pi)^2}\,\left(\,  \frac{1}{180}\,N_0
+ \frac{1}{30}\,N_{1/2} - \frac{1}{10}\,N_{1}\,\right)\,.
\label{abc}
\eeq
The contribution of the Weyl spinor is half of the Dirac one.
We remark that the last two (topological and surface) terms in 
(\ref{divs}) are very important because they contribute to the 
trace anomaly and thus affect the dynamical equations for the 
effective action of gravity. The one-loop effective action has 
the form
\beq
\Ga = S + {\bar \Ga}\,,\,\,\,\,\,\,\,\,\,\,\,
{\bar \Ga} =  {\bar \Ga}^{(1)} + \De S
\label{total}
\eeq
where ${\bar \Ga}^{(1)}$ is naive quantum correction to 
the classical
action $S$ and $\De S$ is a counterterm. 
The anomalous trace of the Energy-Momentum Tensor is
\cite{duff}
\beq
T = <T_\mu^\mu> = - \left[\,w C^2 + b E
 + c {\Box} R\,\right]
\label{anomaly}
\eeq
with the same coefficients $a,b,c$. The Eq. (\ref{anomaly}), 
in turn, gives rise to the equation for the finite part of the 
1-loop correction to the effective action
\beq
\frac{2}{\sqrt{-g}}\,g_{\mu\nu}
\frac{\de {\bar \Ga}}{\de g_{\mu\nu}} 
= wC^2 + bE + c{\Box} R
\label{mainequation}
\eeq
The solution of this equation is straightforward \cite{rei,frts}
and one gets, in terms of 
$\,g_{\mu\nu}={\bar g}_{\mu\nu}\cdot exp[2\si]$,
the following quantum correction to the vacuum action
\footnote{ Recently
there were indications \cite{osborn,deser} that the solution
(\ref{quantum}) does not assume the test based on the calculation
of the three-point function, and hence another solution for 
${\bar \Ga}$
should be found. We emphasize that any solution for the effective
action differs from (\ref{quantum}) by the conformal invariant
functional and this difference is not relevant for the  
cosmological solutions which will be explored here.}. 
$$
{\bar \Ga} = S_c[{\bar g}_{\mu\nu}] + 
\int d^4 x\sqrt{-{\bar g}}\,\{ 
w\si {\bar C}^2 + b\si ({\bar E}-\frac23 {\bar {\Box}}
{\bar R}) + 2b\si{\bar \De}\si -
$$
\beq
- \frac{1}{12}\,(c+\frac23\,b)\,[{\bar R} - 6({\bar \na}\si)^2 - 
6({\bar \Box} \si)]^2)\}\,.
\label{quantum}
\eeq
Here the fiducial metric ${\bar g}_{\mu\nu}$ has fixed determinant,
$\,\De\,$ is conformal invariant self-adjoint operator
$$
\De = {\Box}^2 + 2R^{\mu\nu}\na_\mu\na_\nu
-\frac23\,R{\Box}+\frac13\,(\na^\mu R)\na_\mu
$$ 
and $S_c[g_{\mu\nu}] = S_c[{\bar g}_{\mu\nu}]$ 
is some unknown conformal-invariant 
functional. In general, exact calculation of this functional is
impossible. However, if we are interested in the conformally flat
cosmological solutions, this functional is (fortunately) of no
importance for us. 

\section{Particular inflationary solution}

Our purpose is to implement the quantum correction (\ref{quantum})
into the classical gravity-matter system. For the sake of simplicity
here we consider only the conformally flat metrics,
for which  $S_c[{\bar g}_{\mu\nu}]$ is irrelevant.
The solution (\ref{quantum}) should be added to the classical action
$S_{vacuum}+S_{matter}$. However, since we are interested in the
conformally flat matter, it decouples from the conformally flat 
metric
\footnote{ This fact can be seen, in particular, from the identity
$T^ \mu_\mu=0$
which holds for the Energy-Momentum Tensor of the radiation.}. 
Hence the only nontrivial 
contribution to the dynamical equation for $\si$ comes from the 
vacuum part. The conformal and surface terms 
in the vacuum action (\ref{vacuum}) do not contribute to the 
dynamical equations for the conformal factor.
On the other hand, in order to have correspondence with 
classical gravity one has to add the Einstein-Hilbert term 
to the vacuum action. Then we meet the total action of the form 
\beq
S_{t} = - \frac{1}{\ka^ 2}\int d^4 x \sqrt{- g}\,R
+ {\bar \Ga}[g] + {\rm conformal \,\,invariant\,\,terms}.
\label{act}
\eeq
and, denoting conformal time as $\eta$ and 
taking conventional $\,\si = \ln a(\eta)\,$ we arrive at 
$\,\int d^4 x \sqrt{- g}\,R = 6\int d^4 x \,{{a}^{\prime}}^2\,$. 
Here the prime stands for the derivative with respect to $\,\eta\,$.

Let us remind that $\ka^{-2} = M_{Pl}^2$.
Taking variational derivative of the total action with respect 
to $a(\eta)$ one arrives at the 
equation of motion for $\,a(\eta)$:
\beq 
\frac{a^{\prime\prime\prime\prime}}{a}
-4\frac{{a}^\prime\,{a}^{\prime\prime\prime}}{a^{2}}
-3\,\frac{{a^{\prime\prime}}^2}{a^{2}} 
+2\left(3-\frac{2b}{c}\right) 
\frac{{a^{\prime\prime}}\,{a^\prime}^{2}}{a^{3}} 
+ \frac{4b}{c}\,\frac{{a^\prime}^{4}}{a^{4}}
- \frac{2}{c}\,M_{Pl}^2\,
{a^{\prime\prime}}\,a=0 \,.
\label{eq}
\eeq
It is convenient to rewrite (\ref{eq}) in terms of physical time 
$\,t\,$ where $\,a(\eta)d\eta = dt\,$ so that 
$\,{d}/{d\eta}=a\,{d}/{dt}\,\,$ 
etc, and denote the derivatives with respect to $\,t\,$ by points.
\beq
\label{foe}
a^2\,{\stackrel{....} {a}}
+3\,a\,{\stackrel{.} {a}} \,{\stackrel{...} {a}}  
- \left(5 + \frac{4b}{c}\right)\,{\stackrel{.} {a}}^2\,{\stackrel{..}{a}}
+ a\,{\stackrel{..} {a}}^2 
- \frac{2M_{Pl}^{2}}{c}\left( a^2 \,{\stackrel{..} {a}}
+ a\,{\stackrel{.} {a}}^2\right)\, = \,0\,.
\label{para t}
\eeq
The last equation contains two dimensionless constants $\,b,c\,$
which depend on the particle content of the matter fields and may be 
fixed by choosing particular gauge model. 
Also there is a dimensional constant $\,M_{Pl}\,$, which 
defines the scale. The Eq. (\ref{para t}) 
looks quite complicated and it is difficult to solve it exactly. 
Here we explore a particular exponential solution and  in the next 
section perform the numerical study and physical analysis.
Below we present all the numerical results for the Minimal Standard 
Model and and for the compactified to $D=4$ supergravity (M-theory). 
Indeed, since gravity is treated as purely classical background, 
in the last case one has to disregard all spin-2 and spin-$3/2$ degrees 
of freedom and keep only contributions of the spin-$(0,1/2,1)\,$ sectors. 
 
Let us look for the exponential solution of the 
form $\,a(t)=a_0(t) = A\cdot exp(\la {t})$ where $\,A, \la\,$
are some constants. Substituting this
expression into Eq. (\ref{para t}) one can easily see that it gives 
solution for an arbitrary $\,A\,$ and 
$\,\la = \pm \frac{M_{Pl}}{\sqrt{-b}}$. Hence the 
crucial point is that $\,b\,$ in (\ref{abc}) is negative for any 
particle content of the original theory (\ref{fields}). 
There are always two solutions with opposite signs for $\,\la\,$, and 
the positive one describes the exponentially expanding Universe.
This exponential solution is exactly the one which has been obtained 
in \cite{mamo,star} by the use of the renormalized Energy-Momentum 
Tensor \cite{bunch} (see also \cite{fhh}). Some comment is in 
order. The equation for $\,a(t)\,$ which was used in \cite{star}
looks different from (\ref{para t}). This is because the in \cite{star} 
the $(0,0)$-component of the equation 
$$
R_{\mu\nu} - \frac12\,Rg_{\mu\nu} = 8\pi <T_{\mu\nu}>
$$
has been used while we use the trace of it. Indeed both equations 
contain the same information since the metric has only one degree 
of freedom. On the other hand, in our framework it is clear that the 
introduction of the conformal invariant matter fields (or 
the action for ideal liquid (\ref{liq}) with $p = \frac{\rho}{3}$)
doesn't change the equation, so the above inflationary solution 
is valid for any (conformal) matter content\footnote{In this point
our analysis differs from the one of \cite{fhh}.}. An important 
consequence is that the density perturbations which do not violate the 
constraint $p = \frac{\rho}{3}$ can not destroy the exponential 
behavior of $\,a(t)\,$. On the other hand, any perturbations which
do violate this constraint imply the appearance of the massive 
particles or cosmological constant. In both cases one can not 
consider these perturbations in the framework of the above model 
based on the conformal invariant fields (\ref{fields}) and conformal 
anomaly (\ref{anomaly}). If the initial theory contains some massive 
fields, than the trace $\,<T_{\mu}^{\mu}>\,$ has non-anomalous 
(both classical and quantum) contributions and one has to take them 
into account. Suppose the initial matter fields have masses of the 
order of $\,100\,GeV\,$ that is typical for the Standard Model. Then 
this matter can be successfully considered as (\ref{fields}) at the 
energies a few orders below the Planck scale $\,10^{19}\,GeV\,$ and 
then the Universe is inflating until the typical energy becomes 
comparable with the masses. After that the density perturbations 
violating the constraint $p = \frac{\rho}{3}$ become relevant and they 
provide the successful exit from the inflationary phase and also 
basis for the formation of the galaxies \cite{much}. 
Of course this scheme implies 
that below a Planck mass there is a sufficient gap on the mass scale,
and the heaviest massive particle below $M_{pl}$ is many orders 
lighter. If this condition is satisfied, one can regard all particles 
as massless, so that the constraint $p=\rho/3$ takes place.  
Then the most important questions concerning the exponential solution 
above are: 
\begin{enumerate}
\item whether the inflationary solution is stable under the metric 
perturbations which are consistent with $p = \frac{\rho}{3}$?

\item whether the period of inflation is sufficient to provide 
the necessary rate of expansion?

\item whether the inflation occurs only for the very special 
initial data or it can be achieved in a more general situation?
Indeed the above solution $\,a(t)= A\cdot exp(\la {t})\,$ implies
that $\,{a}^{(n)}(0) = \la^n {a}(0),\,$. Hence this solution 
is only valid for some 1-dimensional line in the 4-dimensional space 
of the initial data. 
\end{enumerate}
 
In order to understand better the sense of the Planck mass in the 
above expressions, one can introduce the dimensionless time 
$\,\tau = M_{Pl}\cdot t\,$. The equation (\ref{para t}) remains almost 
the same, the only change is $\,M_{Pl}\rightarrow 1$. Then, in terms 
of the Planck units we have the following inflationary solution
\beq
\label{is}
a_0(t)  = A\,e^{+{\tau}/{\sqrt{-b}}}
\label{infl}
\eeq

The Eq. (\ref{infl}) can be used to calculate the necessary 
duration of 
inflation. Suppose we want the Universe to expand in $10^n$ times, 
starting from some moment $\tau_0$. According to (\ref{infl}), the
total rate of inflation is
\beq
\frac{a_0(\tau_0 + \De\tau)}{a_0(\tau_0)} =
\exp\,
\left\{\,4\pi\,
\sqrt{\frac{360}{N_t}}
\,\,\De\tau\right\} \,,\,\,\,\,\,\,\,\,\,
N_t=N_0+11\cdot N_{1/2}+62\cdot N_1\,,
\label{ratio} 
\eeq 
and thus we arrive at 
$$
\,\De\tau = \frac{\ln 10}{4\pi}\,\sqrt{ \frac{N_t}{360}}\cdot n\,.
$$ 
For the Minimal Standard Model $\,N_{0}=8,$ 
$N_{\frac{1}{2}}=48,$ $N_{1}=12\,$ and we arrive at
$\,\De\tau \approx 0.345\cdot n\,$. Taking $n=60$ one meets the
necessary period of inflation $\,\De\tau \approx 20.7\,t_{Pl}$ where
$t_{Pl}$ is the Planck time, which serves as a time unit for $\tau$.
Hence the necessary period of inflation is just one order higher
than the Planck time. If one increases the particle content of the
theory $\,\De\tau\,$ becomes longer. In particular, this happens when 
we use SU(5) GUT or the supersymmetric models. One can remind that the 
contributions of different fields to the coefficient $b$ in (\ref{anomaly})
have the same signs.  In general case the expected time of inflation 
can be evaluated as one-two orders longer than the Planck time,
depending on the choice of the model. 

As it was already mentioned above,
it is important to investigate the metric
perturbations around the solution (\ref{infl}). However this 
meets serious difficulty. While taking an arbitrary perturbations
one can not avoid variations of the fiducial metric
${\bar g}_{\mu\nu}$. In this case one has to study this perturbations
on the basis of the full action (\ref{act}) and the 
conformal invariant terms may be as important as the 
anomalous correction which we deal with.
In principle one can mimic the conformal action by the expressions
described in \cite{alter} but we prefer to postpone 
this consideration for some future work. 
Here we consider only the perturbations of the conformal factor.
It proves useful to rewrite the
equation (\ref{para t}) for the $\si (\tau) = \ln a(\tau)$. 
\beq
{\stackrel{....} {\si}}
+ 7 {\stackrel{...} {\si}}{\stackrel{.} {\si}}
+ 4\,\left(3 - \frac{b}{c}\right)\,
{\stackrel{..} {\si}}{{\stackrel{.} {\si}}}^2
+ 4\,{{\stackrel{..} {\si}}}^2
- 4\,\frac{b}{c}\,{{\stackrel{.} {\si}}}^4 
- \frac{2}{c}\,\left(\,2 {{\stackrel{.} {\si}}}^2 
+ {\stackrel{..} {\si}} \right) = 0
\label{logs}
\eeq
and only then perform the perturbations of 
$\si(\tau) = \si_0(\tau) + x(\tau)$ near the inflationary 
solution $\,\si_0(\tau) = \ln A + \frac{1}{\sqrt{-b}}\,\tau\,\,$.
For the sake of simplicity below we put $\,\ln A = 1$. The 
equation for the perturbations have the form
\beq
{\stackrel{....} {x}}
+ \frac{7}{\sqrt{-b}}\, {\stackrel{...} {x}}
- \frac{2}{b}\,\left( 6 - \frac{b}{c}\right)\, {\stackrel{..} {x}}
- \frac{4}{c\sqrt{-b}}\,\, {\stackrel{.} {x}} = 0
\label{perts}
\eeq
(all derivatives with respect to $\tau$) and can be easily explored.
The stability of the inflationary solution under the perturbations 
of conformal factor strongly depends on the particle content 
$N_0,N_{1/2},N_1$ of the initial action (\ref{fields}).
We remark that the stability depends only on the ratio $\,b/c\,$.
With the proportionally increasing number of fields 
$\,N_{0,1/2,1}\,$  
the necessary period of inflation
becomes longer but the (non)stability of solution (\ref{infl})
doesn't change.

Substituting $\,x=e^{r\cdot\tau}\,$ one meets three
different nonzero roots $r_{1,..,3}$. 
We remark that the existence of the zero root manifests the 
arbitrariness of one of the coordinates in the 4-dimensional space 
of initial data (choice of $\ln A$). This can be also seen if 
one makes perturbations in (\ref{para t}) but in the last case 
a bit more sophisticated consideration is required.  
The values of $r_{1,2,3}$ depend on the  particle content 
$\,N_{0,1/2,1}\,$. Taking the values typical 
for the Minimal Standard Model we arrive at 
\beq
  r_{1/2} \approx  - 9.99 \pm 24.71\, i \,,\,\,\, \, \, \, \, \, \,  
\, \, \, \, \, \,\, \, \, \, \, \, r_3 \approx  - 26.66\,.
\label{roots}
\eeq
Thus in the case of Minimal Standard Model all four roots have negative
real parts. This means that the exponential inflation 
is stable under the small perturbations of the 
conformal factor of the metric. The possible sources of 
instability remain in the perturbations of the other degrees 
of freedom or in the density perturbations discusses above.
It is interesting that the stability analysis performed 
(in a different way) in 
\cite{star} gave different result. The source of the
discrepancy is that in \cite{star} the negative sign of $\,c\,$ 
in (\ref{anomaly}) has been chosen. This sign, indeed, corresponds 
to the contributions of vector fields into (\ref{divs}), while
spin-0 and spin-$1/2$ fields contribute with positive sign. 
Here we presented the results for the field content of the 
Standard Model, and will confirm our conclusion using numerical 
methods. 

Let us consider, as a second example, the M-Theory, which is 
nowadays regarded as a possible candidate for the unified model of 
all interactions. In the framework of M-theory one possible realization
is maximal $\,N=1,\,D=11\,$ supergravity. In this case its particle 
content is unique and the gauge group is SO(8). Then the 
spin-$0,1/2,1$ sector of the compactified to $D=4$ theory has $70$ 
real scalars, $28$ vectors and $56$ Maiorana spinors.
Taking $\,N_{(0,1/2,1)} = (70,28,28)\,$ one meets the roots
\beq
  r_{1/2} \approx  - 54.63 \pm 71.39\, i \,,\,\,\, \, \, \, \, \, \,  
\, \, \, \, \, \,\, \, \, \, \, \,r_3 \approx  + 72.63\,,
\label{roots1}
\eeq
indicating to the instability of the exponential inflation.
The question is whether this means instability of inflation at all?
As we shall see in the next section, the answer is exactly 
opposite.

\section{Numerical study and solutions with bounce}

The fourth order equation (\ref{foe}) depends on two parameters $b$ and
$c$, the first being negative for any matter content, and $c$
being positive or negative depending on the particle multiplet.
Numerical integration of this equation depends on
the value and sign of $b$ and $c$ and on the initial conditions.
Among the many possible scenarios, we can identify three main
ones:
\begin{enumerate}
\item Singular expanding Universe, with oscillations between
accelerating and desaccelerating phases;
\item Inflationary expanding Universe;
\item Non singular Universes, presenting a contraction initial phase,
followed by an expanding phase.
\end{enumerate}

The exact inflationary solution (\ref{is}) represents a particular
choice of initial conditions. The perturbative study performed before
indicates that this solution is stable for the particle
content of the Standard Model (see Figures 1,2).
\par
Since it is inconvenient to make a numerical study (and especially
graphical presentation) of the Universe suffering from 60-orders
inflation, we have explored the behavior of logarithmic variable 
$\si(\tau)$ rather then the $a(\tau)$ itself. It turns out that the
inflation 
occurs for a very wide set of initial conditions, even when these
conditions are taken far away from the exact 
exponential solution (\ref{is}). 
In some cases, depending on the multiplet composition,
the stable solutions correspond to some kind of "hiperinflation"
and expansion perform much faster than exponentially.
	
A numerical study of the initial conditions around those representing
the solution (\ref{is}) reveals that this inflationary
solution is "robust": there exist two main scenarios:
\begin{enumerate}
\item An eternal inflationary expansion;
\item A non singular Universe, presenting a bounce for the scale factor.
\end{enumerate}
In the last case, the Universe exhibits an accelerating contraction
followed by an accelerating expansion; just in some special cases
there is a short non inflationary period of contraction (expansion).
These numerical results confirm the analytical one, and extend
the possible scenarios including Universe exhibiting bounce.
Only these two scenarios were found for the Standard Model (Figures 1,2
give typical examples of the plots).

We remark that the 
cosmological scenario exhibiting bounces may be encounter in
many non linear curvature models and in some special scalar-metric
gravity theories. The main distinguishing feature of the scenarios
obtained here is its almost universal
inflationary behavior, while in the other theories there was
just a short period of inflation near the bounce. The continuosly 
expanding and bouncing
solutions obtained here have the number of $e-$fold sufficient to
solve the problems that can be cope with an inflationary phase.

As a second example we consider the multiplet of the M-Theory
described in the previous section. 
In this case, if one performs a slight change of the initial 
conditions around those typical for the exact inflationary
solution, there can be an extremely fast expansion, with singularity 
just a fraction of (Planck) time unite after the initial moment.
Some "typical" plots are presented at the Figures 3,4. 
One can call the behavior shown at Fig. 3 as "hyperinflation",
while the one on Fig. 4 corresponds to some drammatic contraction
of the Universe. Making the time inverse, however, we meet inflation
also in this case. Solutions with the bounce are also possible.

\section{Conclusions and discussions}

We have considered the quantum effects of free radiation on the
background of conformally flat metric and found that this leads,
in a very natural way, to the inflation which must continue only 
a very short time to resolve the problems  
which inflation is supposed to resolve.

It is important to notice that the existence of exponential
solution and the 
velocity of expansion depends only on $\,\int E$-type topological 
counterterm and absolutely doesn't depend on the 
$\,\int {\Box}R\,$ and $\,\int  C^2$-type counterterms (\ref{divs}). 
This may be important, since in the same manner as we have
added $\,\int R\,$ to the vacuum action of the conformal theory, one
could also add local (classical) $\int R^2$-term with an arbitrary 
coefficient which can mixture with the same term in (\ref{quantum}). 
That is why the $\,\int R^2$-based inflation is not safe in general. 
But in the case of (\ref{infl}) everything depends only on the 
nonlocal term in the effective action (\ref{quantum}) 
which does not suffer from this ambiguity. The only physical requirement 
which is important for inflation is a gap on the mass scale between
Planck mass $M_{pl}$ and heaviest massive particle $M_{max}$. 
In the course of inflation the typical energy scale decreases and, when 
is becomes comparable to $M_{max}$, the mass of the particles can be seen,
the violations of $p=\frac{\rho}{3}$ constraint show up and the 
inflation quits due to the perturbations.

The remaining open question is indeed the stability of inflation 
with respect to arbitrary perturbations of the metric. As it was
mentioned above, this question can not be addressed within the 
exact solution (\ref{quantum}) because it is necessary to use some 
approximation (or imitation) for the unknown conformal invariant 
functional $S_c$. Similar situation takes place for the 
unconstrained density perturbations which can occur at small 
energies (or, in other terms, at later period of the evolution 
of the Universe). In this case one has to derive, in some reasonable 
approximation, the effective action of the massive fields and, 
as a best option, also take into account their interactions.
In case of the weakly interacting conformal field one has to 
count the
effect of back reaction of vacuum to the matter fields 
which produces a slow running from the conformal fixed point \cite{consha}. 
In this case one can explore the possibility to have inflationary 
solutions for the induced gravity.
We are going to reconsider these problems elsewhere.
The present letter contains a brief report on the
possibility to have consistent and natural inflation from the 
vacuum effects and some review of its main features. 
\vskip 6mm

\noindent
{\bf Acknowledgments.}
One of the authors (I.Sh) is grateful to J.A. Helayel-Neto for the 
discussion of the theory (\ref{roots1}) and also acknowledges warm 
hospitality at the Departamento de Fisica of the Federal University 
of Juiz de Fora. 
A.M.P. is grateful to UFMG and FAPEMIG (MG-Brazil) for the 
scholarship and support. The work of J.C.F and I.L.Sh. has been 
supported by CNPq.

\newpage

{\bf FIGURE CAPTIONS}
\begin{figure}[htb]
   \begin{center}
    \vskip -3cm
\hspace*{-3cm}
  \vspace*{-1.0cm}
    \epsfxsize=8cm
    \epsffile{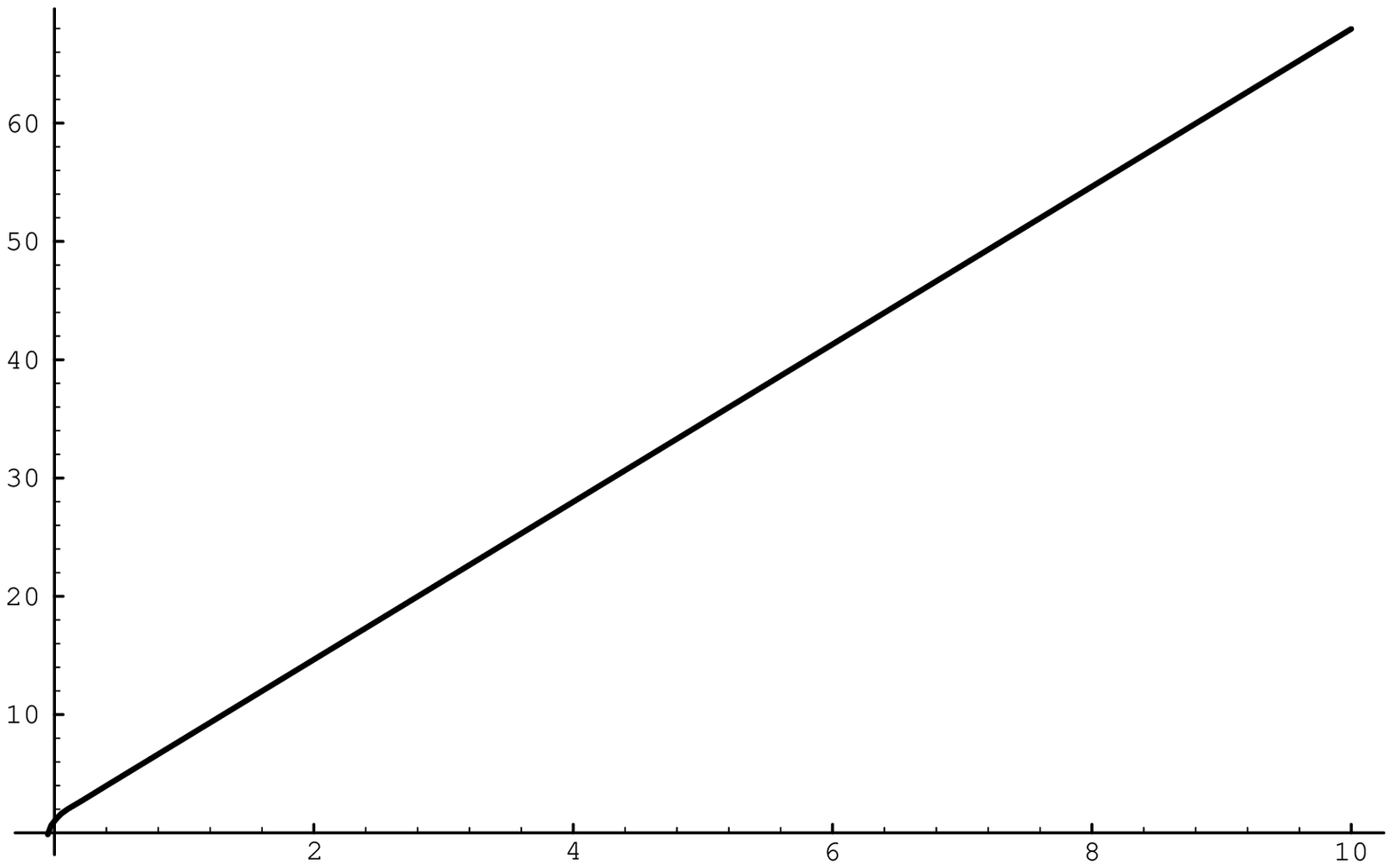}
 \vspace*{-1.0cm}
    \caption{Typical inflationary behavior for the Standard Model case.
Initial data: $\,\,\,\,\,\,\,\,\,\,$ 
$\si(0)=1,\,\, {\stackrel{.} {\si}} (0)= 2m/\sqrt{-b},\,\,  
  {\stackrel{..} {\si}} (0)=0,\,\,
 {\stackrel{...} {\si}} (0)=0$. }
\end{center}
 \end{figure}
\begin{figure}[htb]
   \begin{center}
    \vskip -3cm
\hspace*{-3cm}
  \vspace*{-1.0cm}
    \epsfxsize=8cm
    \epsffile{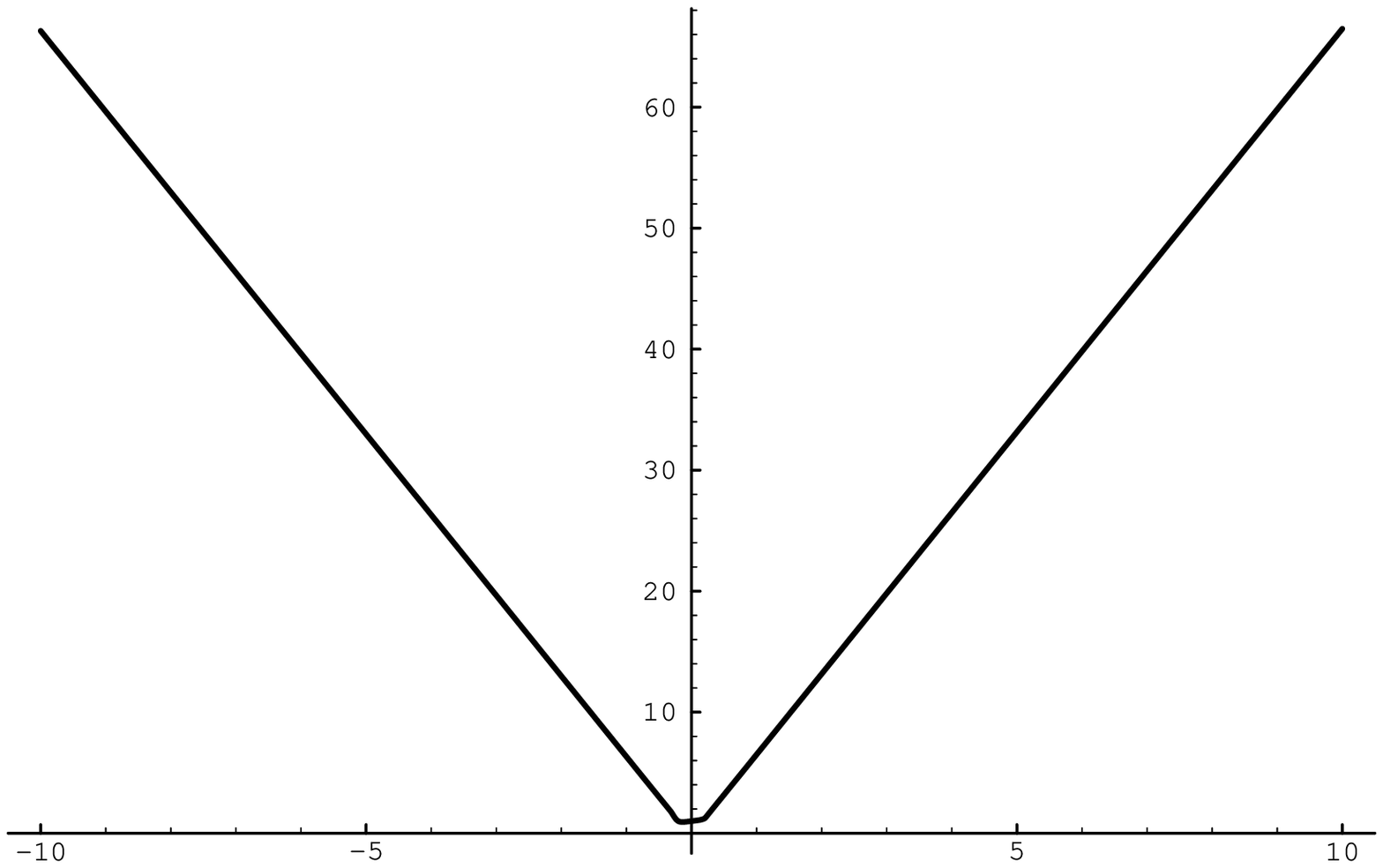}
 \vspace*{-1.0cm}
    \caption{Another 
typical inflationary behavior for the Standard Model case.
Initial data: 
$\si(0)=1, \,\,
{\stackrel{.} {\si}} (0)= 0.1\cdot m/\sqrt{-b},\,\,
{\stackrel{..} {\si}} (0)= 1,\,\, 
{\stackrel{...} {\si}} (0)=0  $. }
\end{center}
 \end{figure}

\begin{figure}[htb]
   \begin{center}
    \vskip -3cm\hspace*{-3cm}
  \vspace*{-1.0cm}
    \epsfxsize=8cm
    \epsffile{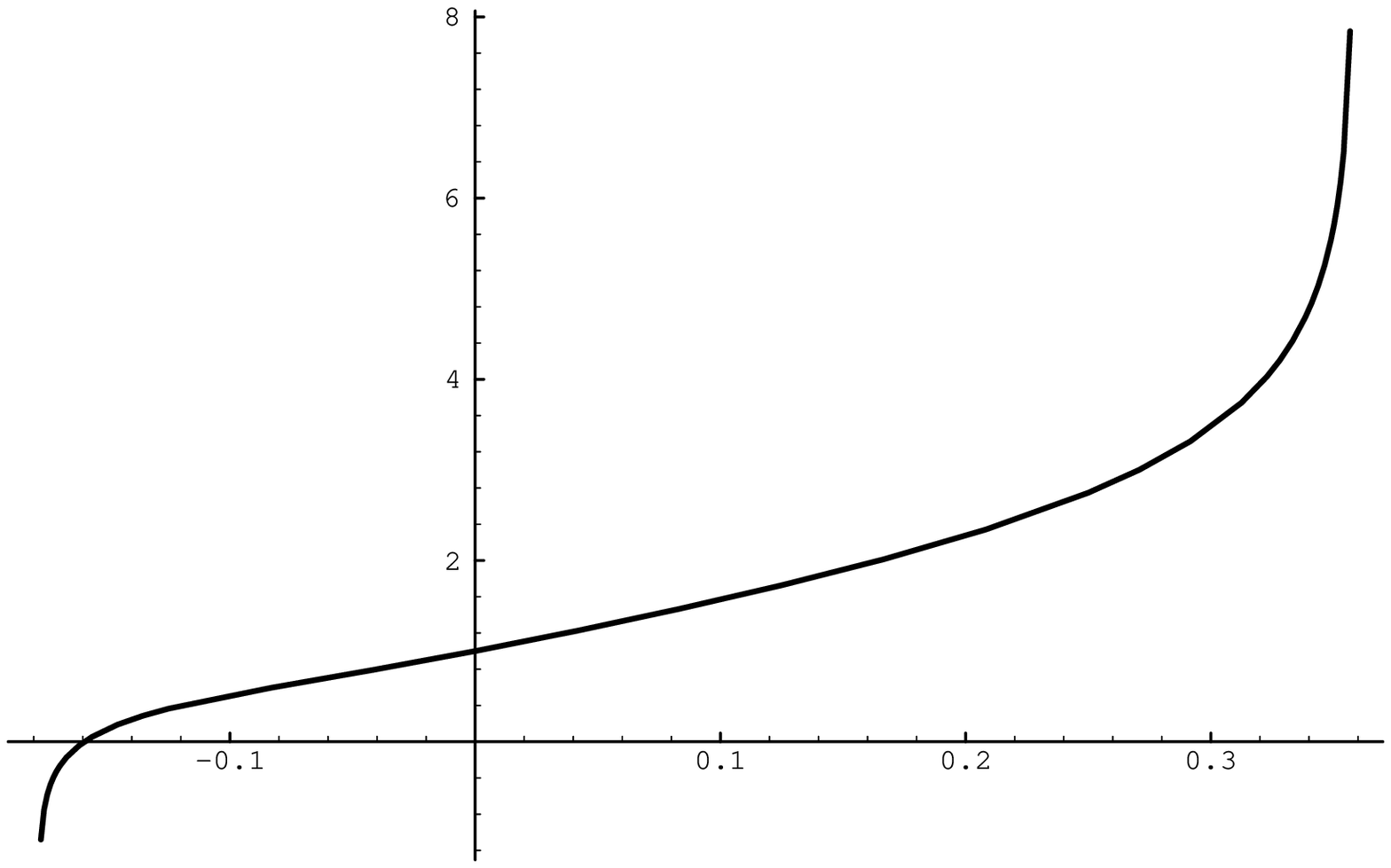}
 \vspace*{-1.0cm}
    \caption{Typical inflationary (hyperinflationary) behavior 
    for the M-theory case.
Initial data: $\si(0)=1, \,\,{\stackrel{.} {\si}} (0)= m/\sqrt{-b},  \,\,
 {\stackrel{..} {\si}} (0)=10,\,\,
 {\stackrel{...} {\si}} (0)=0$. }
\end{center}
 \end{figure}
\begin{figure}[htb]
   \begin{center}
    \vskip -3cm\hspace*{-3cm}
  \vspace*{-1.0cm}
    \epsfxsize=8cm
    \epsffile{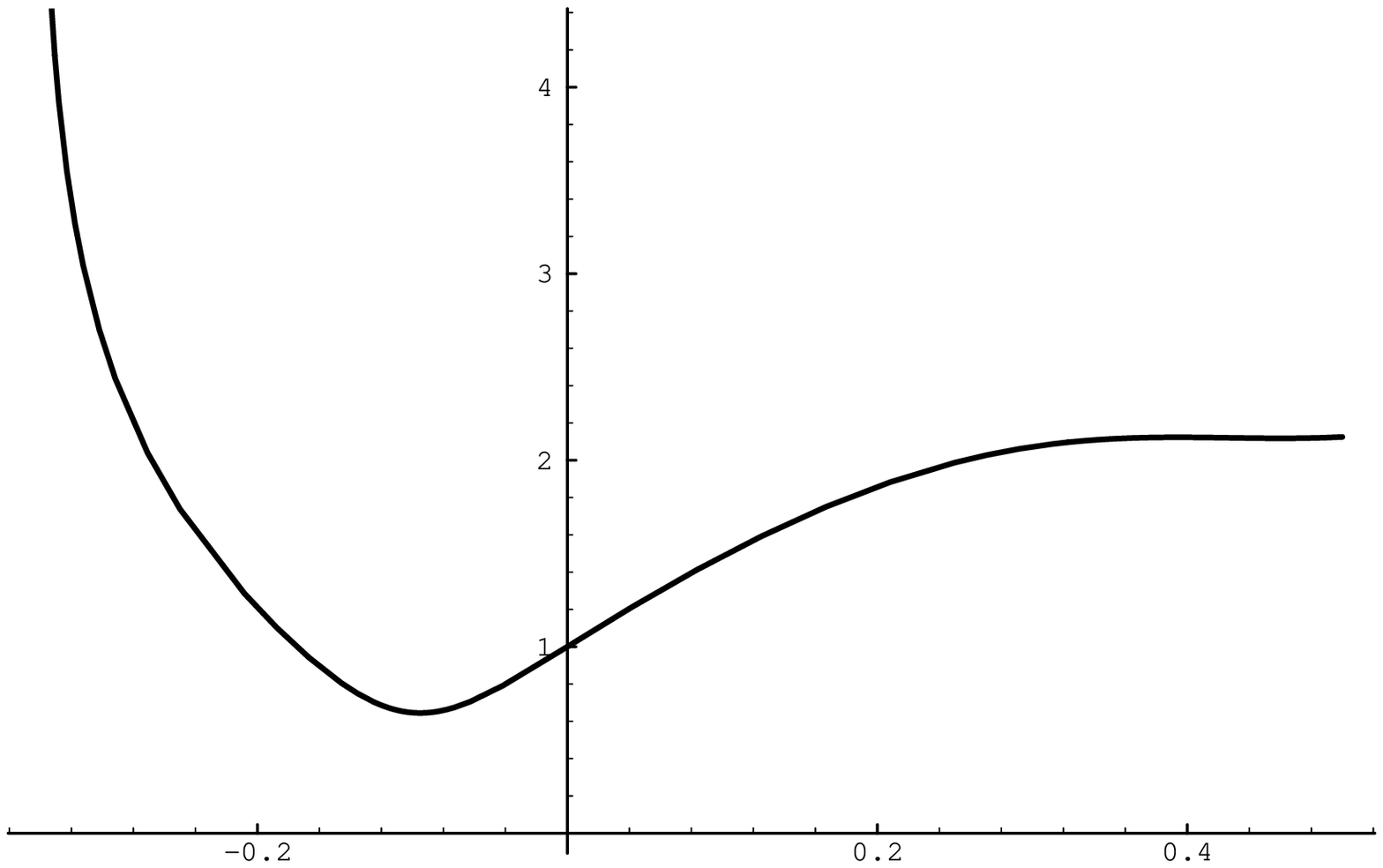}
 \vspace*{-1.0cm}
    \caption{Non-inflationary behavior for the M-theory case.
This kind of scenario with "hyperinflation" is possible 
only for large negative $ {\stackrel{...} {\si}} (0) $.
Initial data corresponding to this plot: 
$\si(0)=1, \,\,{\stackrel{.} {\si}} (0)= m/\sqrt{-b}, \,\, 
{\stackrel{..} {\si}} (0)=0,\,\,
 {\stackrel{...} {\si}} (0) = -400  $. The change of the signs 
in ${\stackrel{...} {\si}} (0)$ and
${\stackrel{.} {\si}} (0)$ is equivalent to the inverse
of time and leads to ``hyperinflation''.}
\end{center}
 \end{figure}

\end{document}